\documentclass[11pt]{article}

\usepackage{amsmath,amsthm,amssymb}

\newcommand{\1}{\,\mathbb{I}}
\newcommand{\bra}[1]{\left\langle{}#1\right|}
\newcommand{\cfield}{\mathbb{C}}
\newcommand{\dff}{\sc}

\newcommand{\ee}{\mathbf{e}}

\newcommand{\hh}{\mathcal{H}}

\newcommand{\hk}{\mathcal{K}}
\newcommand{\hl}{\mathcal{L}}

\newcommand{\ket}[1]{\left| #1\right\rangle}

\newcommand{\kk}{\mathbf{K}}

\newcommand{\lth}{n}

\newcommand{\ppp}[1]{{#1}{#1'}}
\newcommand{\rfield}{\mathbb{R}}
\newcommand{\rrh}{\rho}
\newcommand{\rrp}{\boldsymbol{\rho}}
\newcommand{\twpl}{\boldsymbol{P}}

\DeclareMathOperator{\trc}{Tr}

\newcommand{\cbn}{\cfield{B}_\lth}

\newcommand{\cbp}{\mathfrak{T}}
\newcommand{\bracket}[3]{\bra{#1}#2\ket{#3}}

\newcommand{\braket}[2]{\left\langle{}#1\,\right|\left.#2\right\rangle}
\newcommand{\hhp}{\mathfrak{B}}
\newcommand{\hlp}{\mathfrak{L}}
\newcommand{\ketbra}[2]{\ket{#1}\!\!\bra{#2}}

\newcommand{\mesb}{\,d\mathbf{S}_{\lth}}

\newcommand{\raypr}[1]{\ketbra{#1}{#1}}

\title{
Continuous optimal ensembles II. Reducing the separability
condition to numerical equations}

\author{Rom\`an R. Zapatrin\thanks{Friedmann Lab. for Theoretical
Physics, SPb EF University, Griboyedova 30--32, 191023,
St.Petersburg, Russia; e-mail: zapatrin at rusmuseum.ru}}

\begin{document}

\maketitle

\begin{abstract}
A density operator of a bipartite quantum system is called
robustly separable if it has a neighborhood of separable
operators. Given a bipartite density matrix, its property to be
robustly separable is reduced, using the continuous ensemble
method, to a finite number of numerical equations. The solution of
this system exists for any robustly separable density operator and
provides its representation by a continuous mixture of pure
product states.
\end{abstract}

\section{Introduction}\label{sintro}

Entanglement is treated as a crucial resource for quantum
computation. It plays a central r\^ole in quantum communication
and quantum computation. A considerable effort is being put into
quantifying quantum entanglement. Usually the efforts are focused
on quantifying entanglement itself, that is, describing the
\emph{impossibility} to prepare a state by means of LOCC (local
operations and classical communications). One may, although, go
another way around and try to quantify \emph{separability} rather
than entanglement: this turned out to be applicable for building
combinatorial entanglement patterns for multipartite quantum
systems \cite{myjmo}. Briefly, the contents of this paper is the
following.

I dwell on the case of bipartite quantum systems. A state of such
system is called {\dff separable} if it can be prepared by LOCC,
and {\dff robustly separable} if it has a neighborhood of
separable operators in the space of self adjoint operators. In
terms of density matrices that means that $\rrp$, its density
matrix, can be represented as a mixture of pure product states. I
suggest to replace finite sums of projectors by continuous
distributions on the set of unit vectors. The case of single
particle system is considered in section \ref{scontensemb}, as an
example, the explicit form for the equations for the parameters of
the mixture is obtained.

In section \ref{scontbi} the bipartite case is considered. The
density operators are  represented by distribution on the
Cartesian product of unit spheres in subsystems' spaces. Given a
density operator $\rrp$, we consider it as an element of the space
$\hlp$ of all self adjoint operators in the product space
$\hhp=\hh\otimes\hh$. Then the robust separability of $\rrp$ is
equivalent (see also \cite{coei}) to the solvability of the
following vector equation in $\hlp$:

\[
\nabla
\kk
\;=\;
\rrp
\]

\noindent for the trace functional $\kk$ on $\hlp$ whose form is
obtained in section \ref{scontbi}. When we fix a product basis in
$\hhp$, this equation becomes a system of $\lth^4$ transcendent
equations with respect to $\lth^4$ variables.

\section{Smeared spectral decomposition and
optimal ensembles}\label{scontensemb}

The set of all self-adjoint operators in $\hh=\cfield^\lth$ has a
natural structure of a real space $\rfield^{2\lth}$, in which the
set of all density matrices is a hypersurface, which is the zero
surface of the affine functional $\trc{}X-1$.

Generalizing the fact that any convex combination of density
operators is again a density operator, we represent density
operators as probability distributions on the unit sphere in the
state space $\hh$ of the system.

\paragraph{Technical remark.} Pure states form a projective space
rather than the unit sphere in $\hh$. On the other hand, one may
integrate over any probabilistic space. For technical reasons I
prefer to represent ensembles of pure states by measures on unit
vectors in $\hh$. I use the Umegaki measure on $\cbn$--- the
uniform measure with respect to the action of $U(\lth)$ normalized
so that

\begin{equation}\label{einvarmes}
  \int_{\phi\in\cbn}\limits\;
  \,\mesb
\;=\;
1
\end{equation}

\noindent Similarly, for bipartite case the integration will be carried out
over the Cartesian product of unit spheres in appropriate state
spaces.

\medskip

Now pass to a more detailed account of this issue beginning with
the case of a single quantum system. Let $\hh=\cfield^\lth$ be a
$\lth$-dimensional Hermitian space, let $\rho$ be a density matrix
in $\hh$. We would like to represent the state whose density
operator is $\rrh$ by an ensemble of pure states. Let this
ensemble be continuous with the probability density expressed by a
function $\mu(\phi)$ where $\phi$ ranges over all unit vectors in
$\hh$ (see the technical remark above). The density operator of a
continuous ensemble associated with the measure $\mu(\phi)$ on the
set $\cbn$ of unit vectors in $\hh$ is calculated as the following
(matrix) integral

\begin{equation}\label{e01integral}
  \rrh
  \;=\;
  \int_{\phi\in\cbn}\limits\;
  \mu(\phi)\,
  \raypr{\phi}
  \,\mesb
\end{equation}

\noindent where $\raypr{\phi}$ is the projector onto the vector
$\bra{\phi}$. Effectively the operator integral $\rrh$
\eqref{e01integral} can be calculated by its matrix elements. In
any fixed basis $\{\ket{\ee_i}\}$ in $\hh$, each matrix element
$\rrh_{ij}=\bracket{\ee_i}{\rrh}{\ee_j}$ is the following
numerical integral:

\begin{equation}\label{e01basis}
\rrh_{ij}
\;=\;
  \bracket{\ee_i}{\rrh}{\ee_j}
  \;=\;
  \int_{\phi\in\cbn}\limits\;
  \mu(\phi)\,
  \braket{\ee_i}{\phi}
  \braket{\phi}{\ee_j}
  \,\mesb
\end{equation}

\subsection{Smeared spectral decomposition}\label{ssmspec}

Usual spectral decomposition of a density operator
$\rrh=\sum{}p_k\raypr{\ee_k}$ can be treated as an atomic measure
on $\cbn$ whose density is the appropriate combinations of the
delta functions:

\[
\mu_{\mbox{\scriptsize spec}}(\phi)
\;=\;
\sum{}p_k\,\delta({\phi-\ee_k})
\]

For further purposes a `smeared' version of the spectral
decomposition is needed, begin with some technical setup. Denote
by $p_0$ the smallest eigenvalue of the density matrix $\rrh$
(recall that $p_0>0$ as we restrict ourselves to full-range
density matrices). Let $K$ be an integer such that
$K+1<(1/\lth{}p_0)$, then the density matrix $\rrh$ is represented
as a continuous ensemble with positive density:

\[
\rrh
\;=\;
\sum{}p_k\raypr{\ee_k}
\;=\;
  \int_{\phi\in\cbn}\limits\;
\mu(\phi)
\,
 \raypr{\phi}
  \,\mesb
\]

\noindent with

\begin{equation}\label{esmsing}
\mu(\phi)
\;=\;
\frac{((K+1)\lth)!}{K\cdot(K\lth)!
\,\lth!}
\,\cdot
\sum_k\limits
\left(p_k
\,-
\frac{1}{(K+1)n}
\right)
\left|
\braket{\ee_k}{\phi}
\right|^{\,2Kn}
\end{equation}

\noindent Furthermore, the distribution \eqref{esmsing} tends to
the spectral decomposition of $\rrh$ as $K$ tends to infinity. See
appendix \ref{sderivsmsing} for the proof of formula
\eqref{esmsing}.

\subsection{Optimal entropy ensembles}\label{soptens}

Let us begin with a single particle case. We need to solve the
following variational problem. Given a functional $Q$ on
$L^1(\cbn)$ and a density matrix $\rrh$ in $\hh$, find the
distribution $\mu$ on the set $\cbn$ of unit vectors in $\hh$ such
that

\begin{equation}\label{e03}
\left\lbrace
\begin{array}{l}
  \int_{\phi\in\cbn}\limits\;
  \mu(\phi)\,\raypr{\phi}\mesb
  \;=\;\rrh
   \\
   \qquad
   \\
  Q(\mu)\;\to\; \mbox{extr}
\end{array}
\right.
\end{equation}

\noindent We choose the differential entropy of the distribution
$\mu$ as the functional $Q$:

\begin{equation}\label{e03q}
Q(\mu)
\;=\;
-\int_{\phi\in\cbn}\limits\;
\mu(\phi)\,\ln\mu(\phi)\,\mesb
\end{equation}

\noindent then, according to \eqref{e01basis}, the variational
problem \eqref{e03} takes the form

\[
\left\lbrace
\begin{array}{l}
\int_{\phi\in\cbn}\limits\;
\mu(\phi)
  \braket{\ee_i}{\phi}
  \braket{\phi}{\ee_j}
  \mesb\;=\;\rrh_{ij}
  \\
-\int_{\phi\in\cbn}\limits\;
\mu(\phi)\,\ln\mu(\phi)\,\mesb
\;\to\;
\mbox{extr}
\end{array}\right.
\]

\noindent Solving this variational problem by introducing
Lagrangian multiples $X_{ij}$ we get

\begin{equation}\label{e03a}
-(1+\ln\mu(\phi))
\,-\,
\sum_{ij}
X_{ij}
  \braket{\ee_i}{\phi}
  \braket{\phi}{\ee_j}
\;=\;
0
\end{equation}

\noindent Combining the Lagrange multiples into the operator
$X=\sum_{ij} X_{ij}\ketbra{\ee_j}{\ee_i}$ we have

\begin{equation}\label{e01a}
\mu(\phi)
\;=\;
  e^{-\bracket{\phi}{X}{\phi}-1}
\end{equation}

\noindent and the problem reduces to finding $\mu$ from the
condition

\begin{equation}\label{e04ini}
\int_{\phi\in\cbn}\limits\;
\mu(\phi)
\raypr{\phi}\mesb
\;=\;\rrh
\end{equation}

\noindent which according to \eqref{e01a} and \eqref{e01basis} and
redefining $X:=-\1-X$ can be written as

\begin{equation}\label{e04}
\bracket{\ee_i}{\rrh}{\ee_j}
\;=\;
\int_{\phi\in\cbn}\limits
\;e^{\bracket{\phi}{X}{\phi}}
\braket{\ee_i}{\phi}
\braket{\phi}{\ee_j}\mesb
\end{equation}

\noindent It follows from \eqref{e03a} that the coefficients
$X_{ik}$ can be chosen so that $X_{ik}=\bar{X}_{ki}$. That means
that the problem of finding the optimal ensemble reduces to that
of finding the coefficients of a self-adjoint operator, that is,
to finding $\lth^2$ numbers from $\lth^2$ equations.

\subsection{The explicit form for single particle
case}\label{ssingexpl}

In case of single particle system the equation \eqref{e04} can be
given an explicit form. First note that, given a self-adjoint
operator $X=\sum\,x_k\raypr{\ee_k}$, for any integrable function
$f(x)$ the operator integral

\[
\int_{\cbn}\limits
\,f\bigl(\bracket{\phi}{X}{\phi}\bigr)
\raypr{\phi}
\mesb
\]

\noindent is diagonal in the eigenbasis of $X$ (see the appendix
in \cite{mygibbs} for proof). Therefore, in order to calculate the
integral
\eqref{e04}, we only need its diagonal elements. Calculate first
the functional $\hk$ \cite{coei}:

\begin{equation}\label{edefhk}
\hk(X)
\;=\;
  \int_{\cbn}\limits
\,e^{\bracket{\phi}{X}{\phi}}
\mesb
\end{equation}

\noindent for which the following formula holds

\begin{equation}\label{ekviadet}
\hk(X)
\;=\;
(-1)^\lth\,(\lth-1)!\;
\frac{W_1(x_1,\ldots,x_\lth)}{W(x_1,\ldots,x_\lth)}
\end{equation}

\noindent where $x_1,\ldots,x_\lth$ are the eigenvalues of the
operator $X$,

\[
W(x_1,\ldots,x_\lth)
\;=\;
\mbox{det }
\left\lvert
\begin{array}{cccc}
  1 & 1 & \ldots & 1 \\
  x_1 & x_2 & \ldots & x_\lth \\
  \ldots & \ldots & \ldots & \ldots \\
  x_{1}^{\lth-1} & x_{2}^{\lth-1} & \ldots & x_{\lth}^{\lth-1}
\end{array}
\right\rvert
\]

\noindent  is the Wandermonde determinant, and the matrix $W_1$ is
defined as

\begin{equation}\label{e78det}
  W_1(x_1,\ldots,x_\lth)
\;=\;
\mbox{det }
\left\lvert
\begin{array}{cccc}
  e^{x_1} & e^{x_2} & \ldots & e^{x_\lth} \\
  1 & 1 & \ldots & 1 \\
  x_1 & x_2 & \ldots & x_\lth \\
  \ldots & \ldots & \ldots & \ldots \\
  x_1^{\lth-2} & x_2^{\lth-2} & \ldots & x_\lth^{\lth-2}
\end{array}
\right\rvert
\end{equation}

\medskip

We can then explicitly write down the expressions for the
functional $\hk$

\begin{equation}\label{e77-90}
\hk
\;=\;
  (\lth-1)!
\,
\sum_{k=1}^{\lth}\frac{e^{x_k}}{\prod_{\stackrel{i=1}{i\neq
k}}^{\lth}\,(x_k-x_i)}
\end{equation}

\noindent For the operator \eqref{e04} we have
$\bracket{\ee_j}{\nabla\,\hk}{\ee_j}\;=$

\begin{equation}\label{e77-90a}
\;=\;
  (\lth-1)!
\,\left[
\sum_{\stackrel{k=1}{k\neq j}}^{\lth}
\frac{e^{x_k}}{(x_k-x_j)\prod_{\stackrel{i=1}{i\neq
k}}^{\lth}\limits
\,(x_k-x_i)}
\,+\,
\frac{e^{x_j}}{\prod_{\stackrel{i=1}{i\neq
k}}^{\lth}\limits
\,(x_i-x_j)}
\left(
1-
\sum_{\stackrel{k=1}{k\neq j}}^{\lth}\limits
\frac{1}{x_j-x_k}
\right)
\right]
\end{equation}

\noindent So, in order to obtain the optimal ensemble for the
single particle density matrix $\rrh=\sum\,p_j\,\raypr{\ee_j}$ we
solve the system of $\lth$ equations
$\bracket{\ee_j}{\nabla\,\hk}{\ee_j}=p_j$.

\section{The bipartite case}\label{scontbi}

Let $\rrp$ be a robustly separable density matrix in the product
space $\hh\otimes\hh'$. Then it can be represented (in infinitely
many ways) as a continuous ensemble of pure product states.
Carrying out the same reasoning as in section \ref{soptens} we
conclude (see section \ref{sexistence} for futher details) that
among those continuous ensembles there exists one having the
greatest differential entropy, this will be the ensemble we are
interested in. Like in section \ref{soptens}, formulate the
variational problem. Let $\rrp$ be a density operator in a tensor
product space $\hhp=\hh\otimes\hh'$. The task is to find a
probability density $\mu(\ppp\phi)$ defined on the Cartesian
product $\cbp=\cbn\times\cbn$ of the unit spheres in $\hh,\hh'$,
respectively.

\begin{equation}\label{e03bi}
  \left\lbrace
\begin{array}{l}
  \int_{\ppp\phi\in\cbp}\limits\;
  \mu(\ppp\phi)\,\raypr{\ppp\phi}\ppp\mesb
  \;=\;\rrp
   \\
   \qquad
   \\
  Q(\mu)\;\to\; \mbox{extr}
\end{array}
\right.
\end{equation}

\noindent with

\begin{equation}\label{e03qp}
  Q(\mu)
\;=\;
-\int_{\phi\in\cbn}\limits\;
\mu(\ppp\phi)\,\ln\mu(\ppp\phi)\,\ppp\mesb
\end{equation}

\noindent Proceeding exactly in the same way as with single
particle, we get the following representation:

\begin{equation}\label{erepbi}
  \rrp
\;=\;
  \int_{\ppp\phi\in\cbp}\limits\;
  e^{\bracket{\ppp\phi}{X}{\ppp\phi}}
\,\raypr{\ppp\phi}\ppp\mesb
\end{equation}

\noindent for some self-adjoint operator $X$ in $\hl$.

\subsection{Bipartite separability problem}\label{sbiseparpro}

From a formal point of view, the equation \eqref{erepbi} provides
a solution of bipartite separability problem. In fact, given a
product basis $\{\bra{\ee_{i}\ee_{i'}}\}$ in the product space
$\hhp$, \eqref{erepbi} is a system of $\lth^4$ \textbf{numeric}
equations with respect to $\lth^4$ variables---the matrix elements
of the quadratic form $X$.

\begin{equation}\label{erepbas}
  \rrp_{\ppp{i}\ppp{k}}
\;=\;
  \int_{\ppp\phi\in\cbp}\limits\;
  e^{\bracket{\ppp\phi}{X}{\ppp\phi}}
\,\braket{\ee_{i}\ee_{i'}}{\ppp\phi}
\,\braket{\ppp\phi}{\ee_{k}\ee_{k'}}\ppp\mesb
\end{equation}

If the solution exists, \eqref{erepbas} provides explicitly an
instance of representation of $\rrp$ as a mixture of pure product
states.

Although these equations are transcendental. Even in the simple
case of a single particle in dimension 2, when the operator $X$ is
proved to commute with $\rrh=p_1\raypr{\ee_1}+p_2\raypr{\ee_2}$,
we have two variables $x_1,x_2$ which we have to find from the
following system of equations, which are a special case of
\eqref{e77-90a}:

\begin{equation}\label{esinglex}
\left\lbrace
\begin{array}{ccc}
\frac{e^{x_1}(x_1-x_2-1)+e^{x_2}}{(x_1-x_2)^2} & = & p_1\, \\
\\
\frac{e^{x_2}(x_2-x_1-1)+e^{x_1}}{(x_1-x_2)^2} & = & p_2\,
\end{array}
\right.
\end{equation}

\medskip

\noindent But the essence of the separability problem is the
\emph{existence} of a solution of \eqref{erepbas}.

\subsection{The existence}\label{sexistence}

In this section I show that the existence of a solution of the
equations \eqref{erepbas} for any robustly separable state follows
from the concavity of the functional $Q$ in \eqref{e03bi}. Begin
with a two-dimensional geometrical analogy. Let $S^2=\{(x_1, x_2,
x_3)\mid x_1+x_2+x_3=1;\;x_1, x_2, x_3\ge 0\}$ be a 2-dimensional
simplex in $\rfield^3$ and $\hk$ be a functional on the plane
$\twpl: x_1+x_2+x_3=1$ symmetric with respect to $x_1, x_2, x_3$.
If $\hk$ is \emph{concave}, that is, for any $a, b\in\twpl$ and
any point $q$ lying between $a$ and $b$

\[
\hk(q)
\;\ge\;
\hk(a),
\quad
\hk(q)
\;\ge\;
\hk(b)
\]

\noindent Let $\hk$ be bounded in a domain containing the simplex
$S^2$ and let $l$ be a line intersecting the interior of $S^2$.
Then $\hk$ has a local maximum in the \emph{interior} of $S^2$.

\medskip

Now return to the case of bipartite density matrices. Any such
matrix $\rrp$ can be represented as (an infinite) number of
\emph{pseudomixtures} of pure product states (see, e.g.
\cite{vidaltarrach} for a discussion). Consider the affine space
$\twpl_1$ of all normalized signed measures on the Cartesian
product $\cbp=\cbn\times\cbn$ of unit spheres (it plays the r\^ole
of the plane $\twpl$ in the example above). Given a density
operator $\rrp$, the collection $l_1$ of all pseudomixtures
representing $\rrp$ is an affine submanifold of $\twpl_1$. The set
of all probability measures on $\cbp$ is a simplex $S^\infty$ in
$\twpl_1$.

A density matrix $\rrp$ is robustly separable if and only if $l_1$
intersects the interior. This because $\rrp$ can be represented as
a mixture of products of full-range density matrices in $\hh,\hh'$
each of which can be, in turn, represented by a smeared spectral
decomposition \eqref{esmsing}, that means that there exists a
probability measure representing $\rrp$ which does not vanish
anywhere on $\cbp$

Therefore the solution of the equations \eqref{erepbas} always
exists for any robustly separable $\rrp$ as the functional $\hk$
introduced in \eqref{e03qp} is concave.

\section{Concluding remarks}\label{sconclud}

First sum up the contents of the paper. Given a density operator
$\rrp$ in the product of two Hilbert spaces each dimension $\lth$,
to solve the separability problem is to tell if it has \emph{a}
representation by a mixture of pure product states. In this paper
\emph{the most smeared} distribution on pure product states which
yields $\rrp$ rather than their finite mixture is considered.

\medskip

\noindent The separability problem is reduced to $\lth^4$
numerical equations which are symbolically written as

\[
\nabla\,\hk=\rrp
\]

\noindent whose coordinate form in any product basis in
$\hhp=\hh\otimes\hh$ is \eqref{erepbas}

\[
  \rrp_{\ppp{i}\ppp{k}}
\;=\;
  \int_{\ppp\phi\in\cbp}\limits\;
  e^{\bracket{\ppp\phi}{X}{\ppp\phi}}
\,\braket{\ee_{i}\ee_{i'}}{\ppp\phi}
\,\braket{\ppp\phi}{\ee_{k}\ee_{k'}}\ppp\mesb
\]

\noindent These equations are transcendental even for non-product
case \eqref{esinglex}, but their solution always exists for any
robustly separable density matrix $\rrp$.

\medskip

That is why one can look for asymptotic methods of finding the
solution of the separability problem along these lines. This issue
will be addressed in the next paper on optimal ensembles.

\paragraph{Acknowledgments.} Several crucial issues related to
this research were intensively discussed during the meeting
Glafka-2004 `Iconoclastic Approaches to Quantum Gravity' (15--18
June, 2004, Athens, Greece) supported by QUALCO Technologies
(special thanks to its organizers---Ioannis Raptis and Orestis
Tsakalotos). Much helpful advice from Serguei Krasnikov is highly
appreciated.

The financial support for this research was provided by the
research grant No. 04-06-80215a from RFFI (Russian Basic Research
Foundation).


\newpage

\appendix

\section{Smeared spectral decomposition}\label{sderivsmsing}

The formula \eqref{esmsing} is derived as follows. Let $\bra{\ee}$
be a unit vector in $\hh$, then for any integer $m$ the following
formula holds

\begin{equation}\label{ei07}
  \int_{\phi\in\cbn}\limits\;
\left|
\braket{\ee}{\phi}
\right|^{\,2m}
 \raypr{\phi}
  \,\mesb
\;=\;
\frac{m!\,(\lth-1)!}{\,(m+\lth)!}
\,(\,m\raypr{\ee}+\1)
\end{equation}

\noindent which is a direct consequence of a more general formula
obtained in \cite{mygibbs}. Let $m=K\lth$ for some integer $K$.
Then it follows directly from \eqref{ei07} that

\[
K\lth
\raypr{\ee}+
\1
\;=\;
  \int_{\phi\in\cbn}\limits\;
\frac{(K\lth+\lth)!}{(K\lth)!
\,(\lth-1)!}
\left|
\braket{\ee}{\phi}
\right|^{\,2Kn}
 \raypr{\phi}
  \,\mesb
\]

\noindent Dividing this equation by $\lth(K+1)$ we obtain the
following continuous representation of the projector $\raypr{\ee}$
and the white noise matrix $\Lambda=\1/\lth$:

\begin{equation}\label{e77-32}
\frac{K}{K+1}
\cdot
\raypr{\ee}+
\frac{1}{K+1}
\cdot
\frac{\1}{n}
\;=\;
  \int_{\phi\in\cbn}\limits\;
\frac{((K+1)\lth)!}{(K+1)
\cdot(K\lth)!
\,\lth!}
\left|
\braket{\ee}{\phi}
\right|^{\,2Kn}
 \raypr{\phi}
  \,\mesb
\end{equation}

\noindent The formula \eqref{e77-32} is valid for any eigenvector
$\bra{\ee_k}$ of $\rrh$. Form a convex combination of the lhs of
\eqref{e77-32} with (yet unknown) coefficients $q_k$ and require
it to be $\rrh$:

\[
\sum \,q_k
\left(
\frac{K}{K+1}
\cdot
\raypr{\ee_k}+
\frac{1}{K+1}
\cdot
\frac{\1}{n}
\right)
\;=\;
\sum{}p_k\raypr{\ee_k}
\]

\noindent When $\1$ is replaced by $\sum\raypr{\ee_k}$, the second
summand in the lhs of the above formula takes the form
$\frac{1}{K+1}
\cdot
\frac{1}{n}
\cdot
\sum\,
\raypr{\ee_k}
$, from which we obtain

\[
  q_k\cdot
\frac{K}{K+1}
\,+
\frac{1}{K+1}
\cdot
\frac{1}{n}
\;=\;p_k
\]

\noindent the we get the expression for $q_k$

\begin{equation}\label{epkqk}
  q_k
\;=
\;
\frac{K+1}{K}
\left(p_k
\,-
\frac{1}{(K+1)n}
\right)
\end{equation}

\noindent So, we can form the convex combinations of the
expressions \eqref{e77-32} with the coefficients $q_k$ which
yields us $\rrh$:

\[
\rrh
\;=\;
\sum_k\limits
\frac{K+1}{K}
\left(p_k
\,-
\frac{1}{(K+1)n}
\right)
\cdot
  \int_{\phi\in\cbn}\limits\;
\frac{((K+1)\lth)!}{(K+1)
\cdot(K\lth)!
\,\lth!}
\left|
\braket{\ee_k}{\phi}
\right|^{\,2Kn}
 \raypr{\phi}
  \,\mesb
\]

\noindent Let $p_0$ be the smallest eigenvalue of $\rrh$, then, if
$K+1<(1/np_0)$, all the coefficients $q_k$ in \eqref{epkqk} are
positive.

\section{Proof of the formula \eqref{e77-90}}\label{s77-90}

Given a self-adjoint operator $X$ in $\hh$, consider its
eigenbasis $\{\ee_k\}$ and the integral

\[
F^{\lth}(x_1,\ldots,x_\lth)
\;=\;
\int_{\cbn}\limits
\,f\bigl(\bracket{\phi}{X}{\phi}\bigr)
\mesb
\]

\noindent In order to calculate it denote
$\braket{\phi}{\ee_k}=e^{i\theta}\,r_k$. Then, taking into account
that the integrand does not depend on phases, the above integral
(with respect to the normalized measure) reads

\begin{equation}\label{efnj}
F^{\lth}(x_1,\ldots,x_\lth)
\;=\;
\frac{(n-1)!}{2\pi^{\lth}}
\cdot
(2\pi)^{\lth}
\int_{\sum_{k=1}^{\lth-1}\limits r_{k}^2\le 1}\limits
  f\left(x_\lth+
  \sum_{k=1}^{\lth-1}\limits{}
  (x_k-x_\lth)\,r_{k}^2
  \right)
  \,
\left(
\prod_{k=1}^{\lth-1}
r_k\,dr_k
\right)
\end{equation}

\noindent then, introducing the variables $t_k=r_k^2$, we reduce
the above integral to

\begin{equation}\label{efn}
F^{\lth}(x_1,\ldots,x_\lth)
\;=\;
(n-1)\,!\,
\int_{\sum_{k=1}^{\lth-1}\limits t_{k}\le 1}\limits
  f\left(x_\lth+
  \sum_{k=1}^{\lth-1}\limits{}
  (x_k-x_\lth)\,t_{k}
  \right)
\left(
\prod_{k=1}^{\lth-1}
dt_k
\right)
\end{equation}

\noindent For $f(z)=e^z$ the above equation reads
$\frac{F^{\lth}(x_1,\ldots,x_\lth)}{(n-1)\,!} =\,e^{x_\lth}\!\!
\int_{\sum_{k=1}^{\lth-1}\limits t_{k}\le 1}\limits
  \prod_{k=1}^{\lth-1}\limits{}
  e^{(x_k-x_\lth)\,t_{k}}
dt_k
\;=\;
$

\[
\;=\;
\,e^{x_\lth}\!\!
\int_{\sum_{k=1}^{\lth-2}\limits t_{k}\le 1}\limits
  \prod_{k=1}^{\lth-2}\limits{}
  e^{(x_k-x_\lth)\,t_{k}}
dt_k
\int_0^{1-\sum_{k=1}^{\lth-2}\limits t_{k}}\limits
\,e^{(x_{\lth-1}-x_\lth)\,t_{\lth-1}}
\,dt_{\lth-1}
\;=\;
\]
\[
\;=\;
\frac{e^{x_\lth}}{x_{\lth-1}-x_\lth}\!\!
\int_{\sum_{k=1}^{\lth-2}\limits t_{k}\le 1}\limits
  \prod_{k=1}^{\lth-2}\limits{}
  e^{(x_k-x_\lth)\,t_{k}}
dt_k
\,
\left(e^{(x_{\lth-1}-x_\lth)\,(1-\sum_{k=1}^{\lth-2}t_{k})}
-1
\right)
\;=\;
\]
\[
\;=\;
\frac{e^{x_{\lth-1}}}{x_{\lth-1}-x_\lth}\!\!
\int_{\sum_{k=1}^{\lth-2}\limits t_{k}\le 1}\limits
  \prod_{k=1}^{\lth-2}\limits{}
  e^{(x_k-x_{\lth-1})\,t_{k}}
dt_k
\;-\;
\frac{e^{x_\lth}}{x_{\lth-1}-x_\lth}\!\!
\int_{\sum_{k=1}^{\lth-2}\limits t_{k}\le 1}\limits
  \prod_{k=1}^{\lth-2}\limits{}
  e^{(x_k-x_\lth)\,t_{k}}
dt_k
\]

\noindent and we get the following recurrent relation

\begin{equation}\label{erecf}
F^{\lth}(x_1,\ldots,x_\lth)
\;=\;
\frac{\lth-1}{x_{\lth-1}-x_\lth}
\,
\left(
\vphantom{\int_0^1}
F^{\lth-1}(x_1,\ldots,x_{\lth-1}) -
F^{\lth-1}(x_1,\ldots,x_{\lth-2},x_{\lth})
\right)
\end{equation}

\noindent For $\lth=2$ we have the explicit expression like
\eqref{esinglex}, for higher $\lth$ one can directly verify that
the expression \eqref{e77-90} satisfies the recurrent relation
\eqref{erecf}.


\begin{thebibliography}{99}

\bibitem{vidaltarrach}
G. Vidal, R. Tarrach, \emph{Robustness of entanglement}, Phys.
Rev. \textbf{A59}, 141--155 (1999); eprint quant-ph/9806094

\bibitem{myjmo}
R.R.Zapatrin, \emph{Combinatorial Topology Of Multipartite
Entangled States}, Journal of Modern Optics, \textbf{50}, 891--899
(2003); eprint quant-ph/0207058

\bibitem{mygibbs}
R.R.Zapatrin, \emph{A note on continuous ensemble expansions of
quantum states}; eprint quant-ph/0403105

\bibitem{coei}
R.R.Zapatrin, \emph{ Continuous optimal ensembles I: A geometrical
characterization of robustly separable quantum states}; eprint
quant-ph/0503173

\end{thebibliography}
\end{document}